\documentclass[aps,prd,nofootinbib,11pt]{revtex4}
\usepackage{amsfonts}
\usepackage{txfonts}
\usepackage{graphicx}
\usepackage{dcolumn}
\usepackage{bm}
\usepackage{amssymb}
\usepackage{latexsym}

\newcommand{\be}{\begin{equation}}
\newcommand{\ee}{\end{equation}}
\newcommand{\bq}{\begin{eqnarray}}
\newcommand{\eq}{\end{eqnarray}}

\begin{document}

\title{A exact de Sitter cosmological solution of quadratic gravitation with torsion}

\author{Guo-Ying Qi}
\address{College of physics and electronics, Liaoning Normal University, Dalian,
116029, China,\\
Purple Mountain Observation, Academia Sinica, Nanjing, 210008, China }
\author{Yongxin Guo}
\address{Physics Department, Liaoning University, Shenyang 110036, China}

\begin{abstract}
A exact de Sitter-like cosmological solution of quadratic gravitation with
torsion has been found. In the limit of constant energy and pressure, it
becomes a exact de Sitter spacetime. It exists in a wide class of quadratic
gravity theories and is the same in vacuum for all the models in this class,
no matter how the coefficients of the quadratic terms in the Lagrangian are.
It describes an accelerating universe and gives a cosmological constant
which is of the order of magnitude of the observed value. In vacuum the
universe is a de Sitter spacetime without torsion. When matter presents,
however, the spacetime is equipped with curvature as well as torsion. In
other wards torsion can be generated by the energy-momentum of matter
(energy and pressure).

\end{abstract}

\maketitle

The gravitational effects of torsion have been studied extensively in
astronomy of solar system and cosmology in recent years. In most theories of
gravity involving torsion, the source for torsion is assumed to be the
intrinsic spin of matter. Since the spins of fermions are normally randomly
oriented in macroscopic bodies, the amount of torsion generated by
macroscopic bodies is normally negligible. However, in a recent paper, Mao
et al. [1] argue that this assumption has a logical loophole which can and
should be tested experimentally, and consider non-standard torsion theories
in which torsion can be generated by macroscopic rotating objects. They
point out that there is a class of theories, including the Hayashi-Shirafuji
[2] theory, in which the angular momentum of macroscopic spinning bodies
generates a significant amount of torsion. Following [1], some searches for
new gravitational physics phenomena based on Riemann-Cartan theory of
General Relativity including spacetime torsion have been reported [3].
Starting from the parametrized torsion framework of [1], the motion of test
bodies in the presence of torsion, has been analyzed. On the other hand,
Flanagan and Rosenthal have shown [4] that the Einstein-Hayashi-Shirafuji
Lagrangian propose by Mao et al. has serious defects, while leaving open the
possibility that there may be other viable Lagrangians.

In order to explain observable acceleration of cosmological expansion some
cosmological models in the framework of the Poincare gauge theory of gravity
have been proposed [5]. In these models torsion plays an important role and
the Lagrangian includes some quadratic terms of torsion and curvature. In
some models even the vacuum possesses torsion. As an alternative
gravitational theory quadratic gravity without torsion has been introduced
in both classical and quantum gravity for many years, a recent important
development is in critical gravity [6]. In the gauge approach to gravity
torsion is just as geometrical an entity as curvature. Therefore a naive
quadratic theory of gravity should include torsion naturally [7].
Furthermore, if we employ a de Sitter gauge approach to gravity instead of
the Poincare one we can obtain a rather simple Lagrangian including torsion
and a cosmological term automatically appearing [8, 9]:
\begin{equation}
{\cal L}=\frac 1{l^2}\left( aR_{\mu \nu }{}^{\rho \sigma }R^{\mu \nu
}{}_{\rho \sigma }-bT{}^\mu {}_{\nu \rho }T{}_\mu {}^{\nu \rho }+\frac 12R-%
\frac 3{l^2}\right) e,
\end{equation}
where $e$ is the determinant of the co-tetrad $e^I{}_\mu $, and $l$ denotes
the so called de Sitter length.{\em \ }In four dimensional spacetime the
Gauss-Bonnet term $\sqrt{-g}\left[ R_{\mu \nu \lambda \tau }R^{\mu \nu
\lambda \tau }-4R_{\mu \nu }R^{\mu \nu }+R^2\right] $ is purely topological
and then the Lagrangian can be taken as

\begin{equation}
{\cal L}=\frac 1{l^2}\left( \beta R_{\mu \nu }R^{\mu \nu }+\alpha R^2+\gamma
T{}^\mu {}_{\nu \rho }T{}_\mu {}^{\nu \rho }+\frac 12R-\frac 3{l^2}\right) e.
\end{equation}
For the sake of simplicity we can let $\gamma =0$, then we have a more
simple Lagrangian
\begin{equation}
{\cal L}=\frac 1{l^2}\left( \beta R_{\mu \nu }R^{\mu \nu }+\alpha R^2+\frac 1%
2R-\frac 3{l^2}\right) e.
\end{equation}
In contrast to the ordinary quadratic gravity [6, 10], here the field
variables consist of the tetrad{\em \ } $e_I{}^\mu $ and the spin connection
$\Gamma {}^{IJ}{}_\mu $, the spacetime is characterized by curvature as well
as torsion.

Taking into account the above considerations, the purpose of this paper is
to explore the cosmological effects of torsion in quadratic gravity. We will
obtain a exact solution which describes the accelerating expansion of a
spatial flat universe. The spacetime described by this solution is de
Sitter-like, its structure and evolution depend on distribution and motion
of matter. In the limit of constant $\rho $ and $p$, it becomes a exact de
Sitter spacetime. This solution indicates that in vacuum the torsion
vanishes. When matter is present, however, the torsion does no vanish, which
mean that the energy generates not only curvature but also torsion. It is
interesting to observe that these conclusions are independent of the choice
of $\alpha $ and $\beta $, and then are very important properties shared by
a wide class of quadratic gravity theories described by the Lagrangian (3).
It is the cosmological constant $\Lambda $ that make spacetime de
Sitter(-like) in both vacuum and matter. However, whereas in general
relativity and Poincare gauge theory the cosmological constant is a free
parameter, in de Sitter approach it is a intrinsic property of the spacetime
and can be determined in terms of other quantities [11].

The variational principle yields the field equations for the tetrad{\em \ } $%
e_I{}^\mu $ and the spin connection $\Gamma {}^{IJ}{}_\mu $:

\begin{eqnarray}
&&\frac \beta {2l^2}\left( 2e{}^{I\sigma }R{}^\rho {}_\sigma R{}_{\rho \mu
}+2e^J{}_\rho R{}^{\rho \sigma }{}R{}^I{}_{J\mu \sigma }-e{}^I{}_\mu R_{\rho
\sigma }R^{\rho \sigma }\right) +\frac \alpha {2l^2}\left( 4e^{I\nu
}R{}_{\nu \mu }-e{}^I{}_\mu R\right) R  \nonumber \\
&&+\frac 1{2l^2}\left( e^{I\nu }R{}_{\nu \mu }-\frac 12e{}^I{}_\mu R\right) +%
\frac 3{l^4}e{}^I{}_\mu =E^I{}_\mu ,
\end{eqnarray}
\begin{eqnarray}
&&\frac \beta {l^2}e_J{}^\lambda [e_I{}^\mu \partial _\nu R_\lambda {}^\nu
-e_I{}^\nu \partial _\nu R_\lambda {}^\mu +\left( e_I{}^\nu R_\lambda {}^\mu
-e_I{}^\mu R_\lambda {}^\nu \right) e{}^K{}_\tau \partial _\nu e_K{}^\tau
\nonumber \\
&&+e_I{}^\tau \Gamma ^\nu {}_{\nu \tau }R_\lambda {}^\mu +e_I{}^\nu \Gamma
^\tau {}_{\nu \lambda }R_\tau {}^\mu -e_I{}^\mu \Gamma ^\tau {}_{\nu \lambda
}R_\tau {}^\nu -e_I{}^\tau \Gamma ^\mu {}_{\nu \tau }R_\lambda {}^\nu ]
\nonumber \\
&&+\frac \alpha {l^2}[\left( e_I{}^\nu e_J{}^\tau -e_J{}^\nu e_I{}^\tau
\right) \Gamma ^\mu {}_{\nu \tau }R+\left( e_J{}^\mu e_I{}^\nu -e_I{}^\mu
e_J{}^\nu \right) \left( \Gamma ^\lambda {}_{\lambda \nu }R-\partial _\nu
R\right)  \nonumber \\
&&+\left( e_I{}^\nu e_J{}^\mu -e_I{}^\mu e_J{}^\nu \right) Re{}^K{}_\tau
\partial _\nu e_K{}^\tau ]  \nonumber \\
&&+\frac 1{4l^2}[\left( e_I{}^\nu e_J{}^\tau -e_J{}^\nu e_I{}^\tau \right)
\Gamma ^\mu {}_{\nu \tau }+\left( e_I{}^\nu e_J{}^\mu -e_I{}^\mu e_J{}^\nu
\right) \left( \Gamma ^\lambda {}_{\lambda \nu }+e{}^K{}_\tau \partial _\nu
e_K{}^\tau \right) ]=s_{IJ}{}^\mu .
\end{eqnarray}
where $E^I{}_\mu $ and $s_{IJ}{}^\mu $ are energy- momentum and spin tensors
of the matter source, respectively.. We use the Greek alphabet ($\mu $, $\nu
$, $\rho $, $...=0,1,2,3$) to denote indices related to spacetime, and the
Latin alphabet ($I,J,K,...=0,1,2,3$) to denote algebraic indices, which are
raised and lowered with the Minkowski metric $\eta _{IJ}$ $=$ diag ($%
-1,+1,+1,+1$). That may be, these field equations are rather complicated.
They really look nothing like the familiar, well-analyzed equations of GR.
To help understand the significance of these equations, and to use our
previous experience, we will do a translation of (4, 5) into a certain
effective Riemannian form--transcribing from quantities expressed in terms
of the tetrad $e_I{}^\mu $ and spin connection $\Gamma {}^{IJ}{}_\mu $ into
the ones expressed in terms of the metric $g_{\mu \nu }$ and torsion $%
T^\lambda {}_{\mu \nu }$ (or contortion $K^\lambda {}_{\mu \nu }$). The
affine connection $\Gamma ^\lambda {}_{\mu \nu }$ is related to $e_I{}^\mu $
and $\Gamma {}^J{}_{I\mu }$ by
\begin{eqnarray}
\Gamma ^\lambda {}_{\mu \nu } &=&e_I{}^\lambda \partial _\mu e^I{}_\nu
+e_J{}^\lambda e^I{}_\nu \Gamma {}^J{}_{I\mu }  \nonumber \\
&=&\left\{ _\mu {}^\lambda {}_\nu \right\} +K^\lambda {}_{\mu \nu },
\end{eqnarray}
where $\left\{ _\mu {}^\lambda {}_\nu \right\} $, $K^\lambda {}_{\mu \nu }$
are the Christoffel symbol and the contortion, separately, with
\begin{eqnarray}
K^\lambda {}_{\mu \nu } &=&-\frac 12\left( T^\lambda {}_{\mu \nu }+T_{\mu
\nu }{}^\lambda +T_{\nu \mu }{}^\lambda \right) ,  \nonumber \\
T^\lambda {}_{\mu \nu } &=&e_I{}^\rho T^I{}_{\mu \nu }=\Gamma ^\lambda
{}_{\mu \nu }-\Gamma ^\lambda {}_{\nu \mu }.
\end{eqnarray}
Accordingly the curvature can be represented as
\begin{eqnarray}
R^\rho {}_{\sigma \mu \nu } &=&e_I{}^\rho e^J{}_\sigma R^I{}_{J\mu \nu
}=\partial _\mu \Gamma ^\rho {}_{\sigma \nu }-\partial _\nu \Gamma ^\rho
{}_{\sigma \mu }+\Gamma ^\rho {}_{\lambda \mu }\Gamma ^\lambda {}_{\sigma
\nu }-\Gamma ^\rho {}_{\lambda \nu }\Gamma ^\lambda {}_{\sigma \mu },
\nonumber \\
&=&R_{\left\{ {}\right\} }^\rho {}_{\sigma \mu \nu }+\partial _\mu K^\rho
{}_{\sigma \nu }-\partial _\nu K^\rho {}_{\sigma \mu }+K^\rho {}_{\lambda
\mu }K^\lambda {}_{\sigma \nu }-K^\rho {}_{\lambda \nu }K^\lambda {}_{\sigma
\mu }  \nonumber \\
&&+\left\{ _\lambda {}^\rho {}_\mu \right\} K^\lambda {}_{\sigma \nu
}-\left\{ _\lambda {}^\rho {}_\nu \right\} K^\lambda {}_{\sigma \mu
}+\left\{ _\sigma {}^\lambda {}_\nu \right\} K^\rho {}_{\lambda \mu
}-\left\{ _\sigma {}^\lambda {}_\mu \right\} K^\rho {}_{\lambda \nu },
\end{eqnarray}
where $R_{\left\{ {}\right\} }^\rho {}_{\sigma \mu \nu }=\partial _\mu
\left\{ _\sigma {}^\rho {}_\nu \right\} -\partial _\nu \left\{ _\sigma
{}^\rho {}_\mu \right\} +\left\{ _\lambda {}^\rho {}_\mu \right\} \left\{
_\sigma {}^\lambda {}_\nu \right\} -\left\{ _\lambda {}^\rho {}_\nu \right\}
\left\{ _\sigma {}^\lambda {}_\mu \right\} $ is the curvature of the
Christoffel symbol.

We shall focus on cosmology in our subsequent discussion. For the space flat
Friedmann-Robertson-Walker metric
\begin{equation}
g_{\mu \nu }=\text{diag}\left( -1,a\left( t\right) ^2,a\left( t\right)
^2,a\left( t\right) ^2\right) ,
\end{equation}
we have
\begin{eqnarray}
\left\{ _0{}^0{}_0\right\} &=&0,\left\{ _0{}^0{}_i\right\} =\left\{
_i{}^0{}_0\right\} =0,\left\{ _i{}^0{}_j\right\} =a\stackrel{\cdot }{a}%
\delta _{ij},  \nonumber \\
\left\{ _0{}^i{}_0\right\} &=&0,\left\{ _j{}^i{}_0\right\} =\left\{
_0{}^i{}_j\right\} =\frac{\stackrel{\cdot }{a}}a\delta _j^i,\left\{
_j{}^i{}_k\right\} =0,i,j,k,...=1,2,3.
\end{eqnarray}
The non-vanishing torsion components with holonomic indices are given by two
functions $h$ and $f$ [12]:
\begin{eqnarray}
T_{110} &=&T_{220}=T_{330}=a^2h,  \nonumber \\
T_{123} &=&T_{231}=T_{312}=a^3f,
\end{eqnarray}
and then the non-vanishing contortion components are
\begin{eqnarray}
K^1{}_{10} &=&K^2{}_{20}=K^3{}_{30}=0,  \nonumber \\
K^1{}_{01} &=&K^2{}_{02}=K^3{}_{03}=h,  \nonumber \\
K^0{}_{11} &=&K^0{}_{22}=K^0{}_{22}={}a^2h,  \nonumber \\
K^1{}_{23} &=&K^2{}_{31}=K^3{}_{12}=-\frac 12af,  \nonumber \\
K^1{}_{32} &=&K^2{}_{13}=K^3{}_{21}=\frac 12af.
\end{eqnarray}

The non-vanishing components of the Ricci curvature $R_{\left\{ {}\right\}
}{}_{\mu \nu }$ are
\begin{eqnarray}
R_{\left\{ {}\right\} }{}_{00} &=&-3\stackrel{\cdot }{H}-3\stackrel{\cdot }{h%
}-3H^2-3Hh,  \nonumber \\
R_{\left\{ {}\right\} }{}_{11} &=&a^2\left( \stackrel{\cdot }{H}+3H^2+5Hh+%
\stackrel{\cdot }{h}+2h^2-\frac 12f^2\right) ,
\end{eqnarray}
\begin{equation}
R_{\left\{ {}\right\} }{}=6\stackrel{\cdot }{H}+12H^2+18Hh+6\stackrel{\cdot
}{h}+6h^2-\frac 32f^2,
\end{equation}
where $H=\stackrel{\cdot }{a}\left( t\right) /a\left( t\right) $ is the
Hubble parameter. Using these results and supposing the matter source is a
fluid characterized by the density $\rho $ the pressure $p$ and the spin $%
s_{IJ}{}^\mu $ we obtain four independent equations from (4) and (5):
\begin{eqnarray}
&&\left( \beta +3\alpha \right) [-12\left( \stackrel{\cdot }{H}+\stackrel{%
\cdot }{h}\right) ^2-24\left( \stackrel{\cdot }{H}+\stackrel{\cdot }{h}%
\right) H\left( H+h\right)  \nonumber \\
&&+12h\left( h+2H\right) \left( h+H\right) ^2-6\left( h+H\right)
^2f^2+\allowbreak \frac 34f^4]  \nonumber \\
&&+3H^2+6Hh+3h^2-\frac 34f^2-\frac 6{l^2}-2l^2\rho =0,
\end{eqnarray}
\begin{eqnarray}
&&\left( \beta +3\alpha \right) [-4\left( \stackrel{\cdot }{H}+\stackrel{%
\cdot }{h}\right) ^2-8\left( \stackrel{\cdot }{H}+\stackrel{\cdot }{h}%
\right) \left( H^2+Hh\right)  \nonumber \\
&&+\allowbreak 4h\left( h+2H\right) \left( h+H\right) ^2-2\left( h+H\right)
^2f^2+\frac 14f^4]  \nonumber \\
&&-2\left( \stackrel{\cdot }{H}+\stackrel{\cdot }{h}\right) -3H^2-4Hh-h^2+%
\frac 14f^2+\frac 6{l^2}+2l^2p=0,
\end{eqnarray}
\begin{eqnarray}
&&-2\{\left( \beta +6\alpha \right) \left( \stackrel{\cdot \cdot }{H}+%
\stackrel{\cdot \cdot }{h}\right) +3\left( \beta +4\alpha \right) \left(
hH^2+2H\stackrel{\cdot }{H}+2h\stackrel{\cdot }{H}\right) +\left(
\allowbreak 5\beta +18\alpha \right) \left( H\stackrel{\cdot }{h}+h\stackrel{%
\cdot }{h}+h^2H\right)  \nonumber \\
&&+\left( \beta +3\alpha \right) \left( 2h^3-f\stackrel{\cdot }{f}-\frac 12%
hf^2\right) +\frac 14h\}-2l^2s_{01}{}^1=0,
\end{eqnarray}
\begin{eqnarray}
&&f\{2\left( \beta +6\alpha \right) \left( \stackrel{\cdot }{H}+\stackrel{%
\cdot }{h}\right) +6\left( \beta +\allowbreak 4\alpha \right) H^2  \nonumber
\\
&&+2\left( 5\beta +18\alpha \right) Hh+\left( \beta \ +3\alpha \right)
\left( 4h^2-f^2\right) +\frac 12\}-2l^2s_{12}{}^3=0.
\end{eqnarray}

The equations (15) and (16) lead to
\begin{equation}
\stackrel{\cdot }{H}+\stackrel{\cdot }{h}=-2H^2-3Hh-h^2+\frac 14f^2+\frac 4{%
l^2}+\frac 13l^2\left( 3p+\rho \right) .
\end{equation}
and
\begin{eqnarray}
&&\left( \beta +3\alpha \right) \left[ -\frac{16}{l^4}+\allowbreak \frac 8{%
l^2}\left( h+H\right) ^2+\allowbreak \frac 23l^2\left( h+H\right) ^2\left(
3p+\rho \right) \allowbreak -2\frac{f^2}{l^2}-\frac 16\left(
f^2l^2+16\right) \left( 3p+\rho \right) -\frac 19l^4\left( 3p+\rho \right)
^2\right]  \nonumber \\
&&+\frac 14H^2+\frac 12Hh+\frac 14h^2-\frac 1{16}f^2-\frac 1{2l^2}-\frac 16%
l^2\rho =0.
\end{eqnarray}

The time derivative of (19) gives
\begin{equation}
\stackrel{\cdot \cdot }{H}+\stackrel{\cdot \cdot }{h}=-4H\stackrel{\cdot }{H}%
-3\stackrel{\cdot }{H}h-3H\stackrel{\cdot }{h}-2h\stackrel{\cdot }{h}+\frac 1%
2f\stackrel{\cdot }{f}+\frac 13l^2\left( 3\stackrel{\cdot }{p}+\stackrel{%
\cdot }{\rho }\right) .
\end{equation}

In what follows, we suppose $s_{\mu \nu }{}^\lambda =0$. Substituting (21)
and (19) into (17) we are left with
\begin{eqnarray}
&&\frac 14h-\beta \left( h+4H\right) \left( h+H\right) ^2+\frac 4{l^2}\left(
6\alpha h+3\beta h+2\beta H\right)  \nonumber \\
&&+\frac 13l^2\left( 6\alpha h+3\beta h+2\beta H\right) \left( 3p+\rho
\right) +\frac 13\left( 6\alpha +\beta \right) l^2\left( 3\stackrel{\cdot }{p%
}+\stackrel{\cdot }{\rho }\right) -\frac 12\beta f\stackrel{\cdot }{f}%
\allowbreak +\allowbreak \frac 12\beta l^2\left( h+2H\right)  \nonumber \\
&=&0.
\end{eqnarray}
Substituting (19) into (18) yields
\begin{equation}
f\{-\frac 12f^2\beta +\allowbreak \allowbreak 2\beta \left( h+H\right)
^2+\allowbreak 8\frac{\beta +6\alpha }{l^2}+\frac 12+\allowbreak \frac 23%
l^2\left( \beta +6\alpha \right) \left( 3p+\rho \right) \}=0.
\end{equation}
This means
\[
f=0,
\]
or
\begin{equation}
-\frac 12f^2\beta +\allowbreak \allowbreak 2\beta \left( h+H\right) ^2+\frac %
8{l^2}\left( 6\alpha +\beta \right) +\frac 12+\allowbreak \frac 23l^2\left(
\beta +6\alpha \right) \left( 3p+\rho \right) =0.
\end{equation}
So we have two cases:

1) When
\[
f=0
\]
the equations (20) and (22) read, respectively
\begin{equation}
\left( h+H\right) ^2=\frac{\frac 1{2l^2}+\frac 16l^2\rho +\left( \beta
+3\alpha \right) \left( \frac{16}{l^4}+\frac 83\left( 3p+\rho \right) +\frac %
19l^4\left( 3p+\rho \right) ^2\right) }{\left( \beta +3\alpha \right) \left(
\frac 8{l^2}+\allowbreak \frac 23l^2\left( 3p+\rho \right) \right) +\frac 14}%
,
\end{equation}
and
\begin{eqnarray}
&&\frac 14h-\beta \left( h^3+6Hh^2+9H^2h+4H^3\right) +\frac 4{l^2}\left(
6\alpha h+3\beta h+2\beta H\right)  \nonumber \\
&&+2l^2\left( \alpha h+\allowbreak \frac 12\beta h+\frac 13\beta H\right)
\left( 3p+\rho \right) +\frac 13\left( 6\alpha +\beta \right) l^2\left( 3%
\stackrel{\cdot }{p}+\stackrel{\cdot }{\rho }\right)  \nonumber \\
&=&0.
\end{eqnarray}
They are two algebraic equations of $H$ and $h$ and have the solutions

\begin{eqnarray}
H &=&\frac{\left( \frac{24}{l^2}\alpha +\frac{12}{l^2}\beta +\allowbreak
2l^2\left( \alpha +\frac 12\beta \right) \left( 3p+\rho \right) +\frac 14%
\right) A-\beta A^3+\allowbreak \frac 13\left( 6\alpha +\beta \right)
l^2\left( 3\stackrel{\cdot }{p}+\stackrel{\cdot }{\rho }\right) }{3\beta A^2+%
\frac 4{l^2}\left( 6\alpha +\beta \right) +\frac 13\left( 6\alpha +\beta
\right) l^2\left( 3p+\rho \right) +\frac 14},  \nonumber \\
h &=&A-H,
\end{eqnarray}
and

\begin{eqnarray}
H &=&\frac{-\left( \frac{24}{l^2}\alpha +\frac{12}{l^2}\beta \frac{24}{l^2}%
\alpha +2l^2\left( \alpha +\frac 12\beta \right) \left( 3p+\rho \right) +%
\frac 14\right) A+\beta A^3+\allowbreak \allowbreak \frac 13\left( 6\alpha
+\beta \right) l^2\left( 3\stackrel{\cdot }{p}+\stackrel{\cdot }{\rho }%
\right) }{3\beta A^2+\frac 4{l^2}\left( 6\alpha +\beta \right) +\frac 13%
\left( 6\alpha +\beta \right) l^2\left( 3p+\rho \right) +\frac 14},
\nonumber \\
h &=&-A-H,
\end{eqnarray}
where
\begin{equation}
A=\sqrt{\frac{\frac 1{2l^2}+\frac 16l^2\rho +\left( 3\alpha +\beta \right)
\left( \frac{16}{l^4}+\frac 83\left( 3p+\rho \right) +\frac 19l^4\left(
3p+\rho \right) ^2\right) }{\left( 3\alpha +\beta \right) \left( \frac 8{l^2}%
+\frac 23l^2\left( 3p+\rho \right) \right) +\frac 14}}.
\end{equation}

The equations (27) and (28) imply that the spacetime is de Sitter-like, its
structure and evolution are determined by $\rho $ and $p$ i.e. the
distribution and motion of matter. In the limit of constant $\rho $ and $p$,
it becomes a exact de Sitter spacetime. In addition to $H$, the scalar mode $%
h$ of\ torsion also depends on $\rho $ and $p$, which means that the
energy-momentum of matter generates not only curvature but also torsion of
the spacetime. The above {\em s}olutions are, up to our knowledge, new and
may have some interest for cosmology and astronomy.

2) Following from (24), we have
\begin{equation}
f^2=\allowbreak \allowbreak 4\left( h+H\right) ^2+16\frac{6\alpha +\beta }{%
\beta l^2}+\frac 1\beta +\frac 43\frac{6\alpha +\beta }\beta l^2\left(
3p+\rho \right) .
\end{equation}
Substituting it into (20) gives

\begin{eqnarray}
&&-\frac{48\left( \beta +3\alpha \right) \left( \beta +4\alpha \right) }{%
\beta l^4}-\frac{7\beta +24\alpha }{2\beta l^2}-\frac 1{16\beta }-\frac 16%
l^2\rho -\frac{8\left( \beta +3\alpha \right) \left( \beta +4\alpha \right) }%
\beta \left( 3p+\rho \right)  \nonumber \\
&&-\frac{2\left( \beta +4\alpha \right) }{8\beta }l^2\left( 3p+\rho \right) -%
\frac{\left( \beta +3\alpha \right) \left( \beta +4\alpha \right) }{3\beta }%
l^4\left( 3p+\rho \right) ^2=0.
\end{eqnarray}

Differentiating (30) with respect to time and using (19) we obtain
\[
f\stackrel{\cdot }{f}\allowbreak =4\left( h+H\right) \left( -H\left(
h+H\right) +\frac 1{4\beta }+\frac{8\left( \beta +3\alpha \right) }{\beta l^2%
}+\frac{2\left( \beta +3\alpha \right) }{3\beta }l^2\left( 3p+\rho \right)
\right) +\allowbreak \frac{2\left( \beta +6\alpha \right) }{3\beta }%
l^2\left( 3\stackrel{\cdot }{p}+\stackrel{\cdot }{\rho }\right) .
\]
Using these results the equation (22) becomes
\[
0=0.
\]
Therefore, there is no definite solution in this case.

In vacuum
\[
p=\rho =0
\]
(27), (28), and (29) give
\[
A=\frac{\sqrt{2}}l,
\]

\begin{equation}
H=\frac{\sqrt{2}}l,h=0,
\end{equation}
and

\begin{equation}
H=-\frac{\sqrt{2}}l,h=0.
\end{equation}

Let us consider a de Sitter spacetime with $l=l_P$, where $l_P$ is the
Planck length, for which the corresponding cosmological term is
\[
\Lambda =\frac 3{l_P^2}.
\]
We obtain
\[
\Lambda =\frac 32H^2,
\]
which agrees with the result predicted by the quantum gravity approach of de
Sitter relativity [11]. Using the value $H=H_0=75(Km/s)/Mpc$, the
cosmological constant is found to be
\[
\Lambda \simeq 10^{-56}cm^{-2},
\]
which is of the order of magnitude of the observed value [13]. It is worth
notice that the solutions (32) and (33) are independent of the choice of $%
\alpha $ and $\beta $. In other wards, they are the same in all the theories
described by the Lagrangian with arbitrary values of $\alpha $ and $\beta $.

The solutions (32) and (33) mean that vacuum is a de Sitter spacetime
without torsion, whereas the solutions (27) and (28) indicate that when
matter presents, torsion does not vanish. It is the energy $\rho $ and the
pressure $p$ of matter that generate torsion.

In this letter, we have studied the cosmology of a class of quadratic
gravity with torsion described by the Lagrangian (3). We have derived the
basic equations (15-18) and obtained a exact solution (27) and (28). This
solution can be used to explain the observed acceleration of the
cosmological expansion and indicate the macroscopic origin of torsion.

Our results motivate further work. For example, it would be interesting to
address the question of how torsion influence the motion of a particle in
gravitational field on both the{\em \ }theoretical and experimental sides.


\begin{references}
\bibitem{}  Y. Mao, M. Tegmark, A. H. Guth, and S. Cabi, Phys. Rev. D76,
104029 (2007), arXiv:gr-qc/0608121.{\bf \ }

\bibitem{}  K. Hayashi, T. Shirafuji, Phys. Rev. D 19, 3524 (1979).

\bibitem{}  R. March, G. Bellettini, R. Tauraso, and S. Dell'Agnello, Phys.
Rev. D83,104008 (2011), arXiv:1101.2789; arXiv:1101.2791.

\bibitem{}  E. E. Flanagan, E. Rosenthal, Phys. Rev. D75, 124016 (2007){\bf %
, }arXiv:0704.1447.

\bibitem{}  K.-F. Shie, J. M. Nester and H.-J. Yo, Phys. Rev. D78, 023522
(2008), arXiv:0805.3834; H. Chen, F.-H. Ho, J. M. Nester, C.-H. Wang, H.-J.
Yo, JCAP 0910, 027(2009), arXiv:0908.3323; P. Baekler, F. W. Hehl, J. M.
Nester, Phys.Rev.D83, 024001, 2011, arXiv:1009.5112 ; P. Baekler and F. W.
Hehl, arXiv:1105.3504; X.-Z. Li, C.-B. Sun, P. Xi, Phys.Rev.D79,
027301,2009, arXiv:0903.3088,;X.-C. Ao, X.-Z. Li, P. Xi, Phys. Lett. B694,
186 (2010), arXiv:1010.4117. A.V. Minkevich, A. S. Garkun and V. I. Kudin,
Class. Quantum Grav. 24, 5835 (2007), arXiv:0706.1157; A. V. Minkevich,
Phys. Lett. B678, 423 (2009), arXiv:0902.2860.

\bibitem{}  H. Lu and C.N. Pope, Phys. Rev. Lett. 106, 181302 (2011),
arXiv:1101.1971 [hep-th]; S. Deser, H. Liu, H. Lu, C.N. Pope, T. C. Sisman
and B. Tekin, Phys. Rev. D 83, 061502 (2011), arXiv:1101.4009[hep-th]; H.
Lu, Y. Pang and C.N. Pope, arXiv:1106.4657 [hep-th]; Y.-X. Chen, H. Lu and
K.-N. Shao, arXiv:1108.5184 [hep-th].

\bibitem{}  I. L. Shapiro, Phys. Rept. 357, 113 (2002), arXiv:hep-th/0103093.

\bibitem{}  G. Chee, Phys. Rev. D54, 6552 (1996).

\bibitem{}  D. K. Wise, Class. Quantum Grav. 27,155010, (2010),
arXiv:gr-qc/0611154; A. Randono, Class. Quantum Grav. 27, 105008, (2010),
arXiv:0909.5435; G. W. Gibbons and S. Gielen, Class. Quantum Grav. 26,
135005 (2009), arXiv:0902.2001.

\bibitem{}  S. Deser and B. Tekin, Phys.Rev. Lett. 89, 101101 (2002); J. D.
Barrow and S. Hervik Phys. Rev. {\bf D 74},124017 (2006).

\bibitem{}  R. Aldrovandi and J. G. Pereira, arXiv:0711.2274 [gr-qc].

\bibitem{}  M. Tsamparlis, Phys. Lett. 75A, 27 (1979); H. F. M. Goenner and
F. Muller-Hoissen, Class. Quant. Grav. 1, 651 (1984).

\bibitem{}  A. G. Riess et al, Ap. J. 116, 1009 (1998); S. Perlmutter et al,
Ap. J. 517, 565 (1999); P. de Bernardis et al, Nature 404, 955 (2000); S.
Hanany et al, Ap. J. Letters 545, 5 (2000).
\end{references}
\end{document}